\documentclass[journal]{IEEEtran}
\usepackage{graphicx}
\usepackage{amsmath}

\usepackage{stfloats}%gongshi
\ifCLASSINFOpdf
\else
\fi
\begin{document}
\title{Intercore spontaneous Raman scattering impact on quantum key distribution in multicore fiber}

\author{Chun~Cai,
        Yongmei~Sun,
        and~Yuefeng~Ji,~\IEEEmembership{Senoir Member,~IEEE}% <-this % stops a space
\thanks{Yongmei Sun is with the The State Key Laboratory of Information Photonics and Optical Communications, Beijing University of Posts and Telecommunications, Haidian District, Beijing, 100876, China
e-mail: ymsun@bupt.edu.cn.}% <-this % stops a space
\thanks{Chun Cai and Yuefeng Ji are with Beijing University of Posts and Telecommunications.}% <-this % stops a space
%\thanks{This work was supported in part by Fund of National Natural Science Foundation of China under Grant 61831003, 61871051 and State Key Laboratory of Information Photonics and Optical Communications (Beijing University of Posts and Telecommunications), P. R. China (Grant No. IPOC2017ZZ04).}
}



\markboth{Journal of \LaTeX\ Class Files,~Vol.~14, No.~8, August~2015}%
{Shell \MakeLowercase{\textit{et al.}}: Bare Demo of IEEEtran.cls for IEEE Journals}




\maketitle

\begin{abstract}
We propose a model, for the first time, to quantitatively estimate the intercore spontaneous Raman scattering (ICSRS) in multicore fiber (MCF) based on the characterization of intercore crosstalk (ICXT). Also, we show the propertis of ICSRS through numerical simulations. Then the impact of ICSRS on quantum key distribution (QKD) is evaluated. It is revealed that both the forward-ICSRS and backward-ICSRS will reduce the maximum transmisson distance and the forward-ICSRS will reduce more. However, over the range of metropolitan area networks, quantum signals are affected only when the powers of classical signals are very high in dense wavelength division multiplexing system. Finally, the spontaneous Raman scattering (SRS) generated in single core fiber and the ICSRS generated in MCF are compared. The power variety trend with the transmission distance of SRS is similar to that of ICSRS, though there are subtle differences.
\end{abstract}


% \fi
\IEEEpeerreviewmaketitle

\section{Introduction}
\IEEEPARstart{Q}{uantum} key distribution (QKD) allows remote parties to generate secure keys based on the laws of quantum physics \cite{Bennet1984Quantum,Gisin2001Quantum}. It enables information-theoretic communication security and could revolutionize the way in which information exchange is protected in the future \cite{Shor2000Simple}. In the last few decades, many efforts have been made to improve the communication range and secure key rate (SKR) of QKD. Also, lots of progresses have been achieved, such as, the transmission distance of measurement-device-independent QKD in an ultra-low-loss fiber can be as long as 404 km \cite{yin2016measurement}, and high-speed QKD systems with Mbps SKR have been achieved \cite{dynes2016ultra,yuan201810}.

Another trend of practical application of QKD is to integrate it with the classical optical communication. The first scheme of simultaneously transmitting QKD with classical signals was introduced by Townsend in 1997 \cite{townsend1997simultaneous}. The O-band (1260 nm–1360 nm) was chosen for the quantum channel to reduce impairment from the classical channels \cite{townsend1997simultaneous, nweke2005experimental, chapuran2009optical}, which are usually located at C-band (1530 nm–1565 nm). However, the O-band introduces more losses to the faint quantum signal than the C-band. Subsequently, C-band is used to transmit quantum signal in many schemes \cite{eraerds2010quantum, patel2014quantum}. In recent years, QKD is multipelxed with terabit classical data and transmitted over long distances \cite{wang2017long}. Also, QKD is integrated with 3.6 Tbps classical data in a commercial backbone network in 2018 \cite{mao2018integrating}.

QKD is integrated with classical signals in single-mode single-core fiber in the above schemes. The biggest challenge for the integration is the spontaneous Raman scattering (SRS) genarated by classical signal \cite{eraerds2010quantum, peters2009dense}. The spontaneous process converts photons from classical channel into a broad band of wavelengths. It leads to a significant wavelength shift of about 200 nm. Also, the power of the SRS is large enough to affect QKD which makes it the main impairment source to the QKD. Many methods have been proposed to relieve the impact of SRS on QKD, such as spectral-filtering \cite{frohlich2017long, sun2018experimental}, temporal-filtering \cite{patel2012coexistence} and wavelength assignment \cite{bahrani2018wavelength, niu2018optimized}.

Optical networks play an increasingly important role in our lives \cite{ji2018towards} and the data traffic demand in access and backbone networks has been increased exponentially \cite{yu2013transmission}. However, the capacity of existing standard single-core single-mode fiber may no longer satisfy the growing capacity demand and is approaching its fundamental limit around 100 Tbps owing to the limitation of amplifier bandwidth, nonlinear noise, and fiber fuse phenomenon \cite{sakaguchi2013305}. In order to further increase the fiber capacity, space division multiplexing has been proposed and attracted intensive research efforts as a method to solve the capacity saturation of conventional single-mode single-core fiber \cite{richardson2013space, winzer2014making}. Multi-core fiber (MCF) which incorporates multiple separate cores in a single fiber is an effective approach to realize space division multiplexing \cite{saitoh2016multicore}.

Several schemes have been proposed to integrate QKD and classical signals in MCF \cite{dynes2016quantum, lin2019spontaneous, 8535406, eriksson2019crosstalk, cai2019experimental}. The SRS is studied in \cite{dynes2016quantum} and \cite{lin2019spontaneous} experimentally. SRS effect in MCF includes SRS which is generated in the same core by classical signal and intercore SRS (ICSRS) which is generated in another core by classical signal. In \cite{dynes2016quantum}, authors demonstrate that QKD can coexist with classical signals in their 53-km 7-core MCF in the presence of ICSRS, while showing negligible degradation in performance. Based on experimental values, they perform simulations highlighting that classical data bandwidths beyond 1 Tbps can be supported with QKD in their MCF. In \cite{lin2019spontaneous}, the ICSRS and SRS in trench assisted MCF and un-trenched MCF are experimentally measured.

However, no accurate mathematic model has been proposed to evaluate the magnitude of ICSRS and the impact of ICSRS on QKD in MCF. To this end, an analytical expression of the ICSRS, including forward-ICSRS and backward-ICSRS, is derived for the first time. This mathematic model is based on the effect of intercore crosstalk (ICXT) in MCF. Then the properties of ICSRS are analyzed through numerical simulations. It is revealed that both the power of forward-ICSRS and that of backward-ICSRS are dependent on the power coupling coefficient between adjacent cores. The power of the forward-ICSRS reaches a maximum value with the transmission distance and then it starts to decline, while that of the backward-ICSRS saturates and does not decrease with distance. Then the impact of ICSRS on QKD is evaluated. It shows that both forward-ICSRS and backward-ICSRS will reduce the maximum transmission distance of QKD and backward-ICSRS will reduce more. Over the range of metropolitan area networks, the quantum signal is affected by ICSRS only when the power of source classical signal is very high in dense wavelength division multiplexing system. Finally, the SRS generated in single core fiber and the ICSRS generated in MCF are compared. The power variety trend with the transmission distance of SRS is similar to that of ICSRS, though there are subtle differences.

\section{Scenario for integrating QKD with classical signals in MCF}
\label{architecture}

\begin{figure}[!t]
\centerline{\includegraphics[width=\columnwidth]{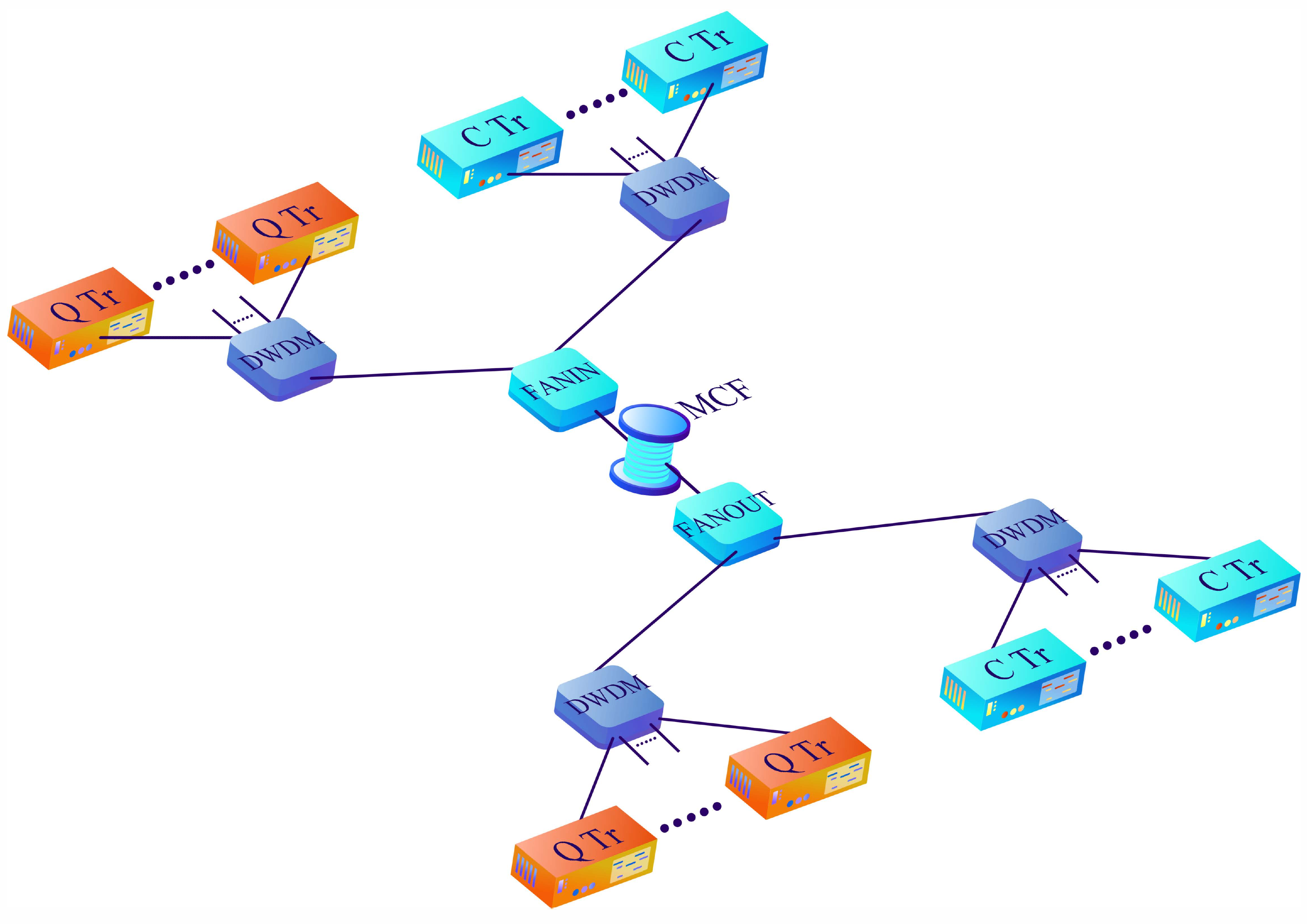}}
\caption{Q Tr: QKD tranceiver; C Tr: classical communication tranceiver; DWDM: dense wavelength division multiplexing module; MCF: multicore fiber.}
\label{fig:architecture}
\end{figure}

Fig.~\ref{fig:architecture} shows the application scenario that will be considered in this paper. Classical signals and quantum signals are multiplxed together in one MCF. Each core of the MCF transmits only quantum signals or classical signals. The transmission direction of quantum signals and classical signals is arbitrary. Under this core assignment scheme, ICXT generated by classical signals is the the main impairment to QKD. ICXT is the power coupling between different cores and its power is mainly concentrate on the spectral peak around the frequency of the source signal. The power of the ICXT (typically -30- -60 dBm) usually higher than that of quantum signal (typically lower than -80 dBm). Thus the frequencies in occupation of classical signals can not be used for quantum signals. Therefore, in this paper, quantum signals and classical signals are assigned in different wavebands in C-band, in which case ICXT can be eliminated by filter easily. However, SRS (more precisely ICSRS in this scenario) cannot be eliminated completely since its bandwidth covers 200 nm \cite{eraerds2010quantum}.

\section{Derivation of the ICSRS equations}
\label{derivation}

\begin{figure}[!t]
\centerline{\includegraphics[width=\columnwidth]{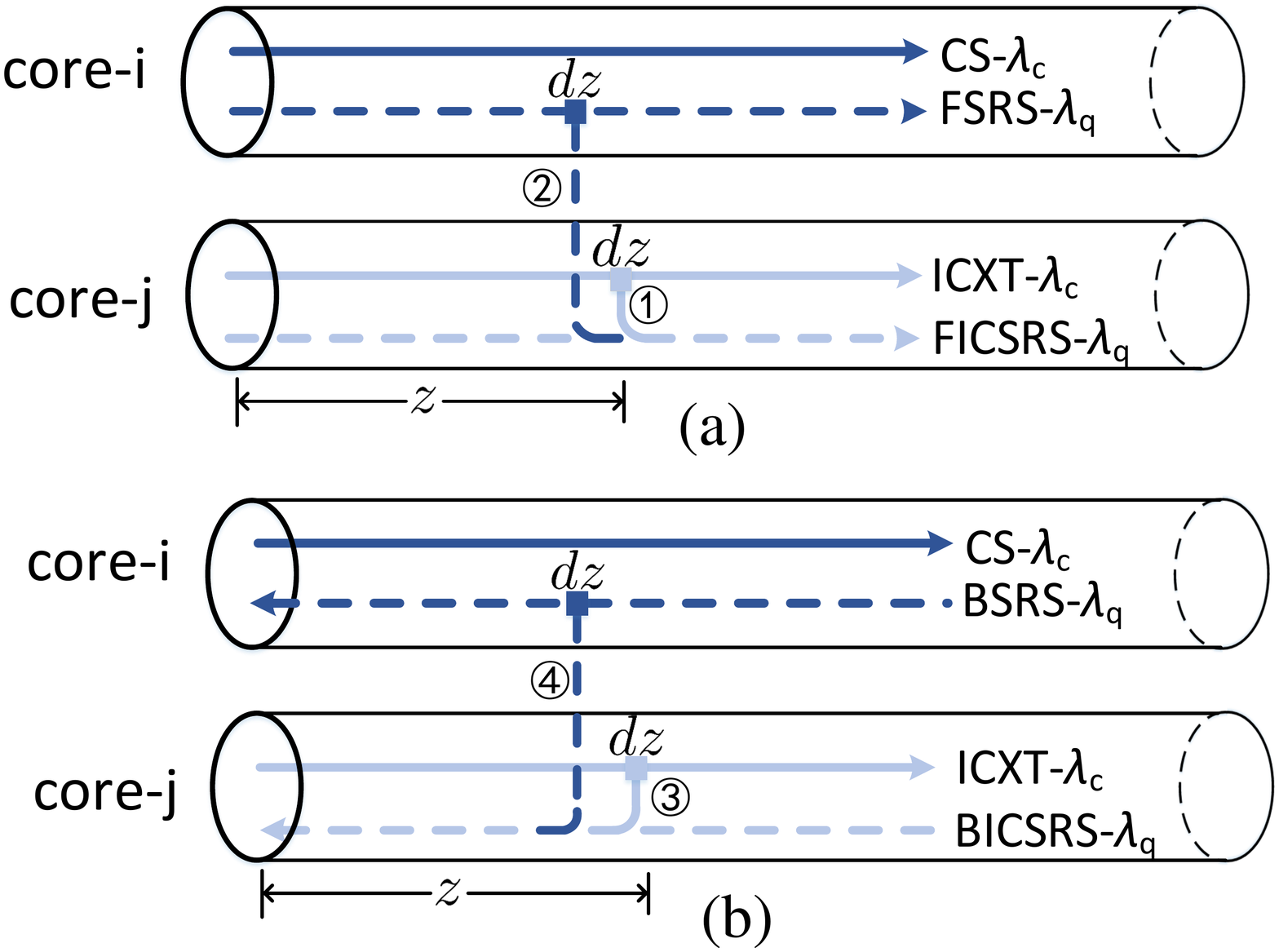}}
\caption{(a) Forward-ICSRS. (b) Backward-ICSRS. (CS: classical signal; FSRS: forward spontaneous Raman scattering; ICXT: intercore crosstalk; FICSRS: forward intercore spontaneous Raman scattering; $\lambda_{c}$: the wavelength of classical signal; $\lambda_{q}$: the wavelength of quantum signal.)}
\label{fig:gstd}
\end{figure}

To quantify the effect of ICSRS on QKD, we first model the generation power of it. The ICSRS includes forward-ICSRS and backward-ICSRS. When the classical signal and the quantum signal are transmitted in the same (co-propagation) or opposite (counter-propagation) direction, forward-ICSRS or backward-ICSRS noise will be introduced to QKD. As shown in Fig.~\ref{fig:gstd}(a), the total forward-ICSRS power is composed of two parts, which is similar to the Rayleigh scattering \cite{ye2016design, sano2014crosstalk}. The classical signal in core-i is the source signal to generate ICSRS. It will generate forward-SRS in core-i at the wavelength of $\lambda_{q}$. Also, it will generate ICXT in core-j at the wavelength of $\lambda_{c}$. Then, this two parts will contribute to ICSRS. The first one is the forward-SRS process of the ICXT, corresponding to the process \textcircled{1} in Fig.~\ref{fig:gstd}(a). The second one is the ICXT process of the forward-SRS, corresponding to the process \textcircled{2}. Thus the power of the forward-ICSRS can be represented as  
\begin{equation}
\begin{split}
\label{eq:FICSRS}
P_{FICSRS}=P_{ICXT-FSRS}+P_{FSRS-ICXT},
\end{split}
\end{equation}
where $P_{ICXT-FSRS}$, $P_{FSRS-ICXT}$ are the power generated by process \textcircled{1}, \textcircled{2}, respectively.

Firstly, we will derive the power generated by process \textcircled{1}. As shown in Fig.~\ref{fig:gstd}(a), the SRS power in a short segment ($dz$) is proportional to the source power \cite{kawahara2011effect}, which can be expressed as 
\begin{equation}
\begin{split}
\label{eq:dP}
dP_{ICXT-FSRS}(z)=\eta P_{ICXT}(z)dz,
\end{split}
\end{equation}
where $\eta$ is the Raman efficiency, $dz$ is the length of one segment. $P_{ICXT}(z)$ is the power of ICXT at $z$ which is described by \cite{ye2016design, sano2014crosstalk}
\begin{equation}
\begin{split}
\label{eq:PICXT}
P_{ICXT}(z)=P_{0}exp(-h_{ij}z)sinh(h_{ij}z)exp(-\alpha_{c}z),
\end{split}
\end{equation}
where $P_{0}$, $h_{ij}$ and $\alpha_{c}$ are the power of classical signal at the input, power coupling coefficient between two adjacent cores (core-i annd core-j in Fig.~\ref{fig:gstd}) and the attenuation coefficient of the classical channel in linear scale, respectively.

When the power describe in Eq. \ref{eq:dP} is at the output of core-j, the power will turn to
\begin{equation}
\begin{split}
\label{eq:dPL}
dP^{L}_{ICXT-FSRS}(z)=dP_{ICXT-FSRS}exp[-\alpha_{q}(L-z)],
\end{split}
\end{equation}
where $\alpha_{q}$ and $L$ are the attenuation coefficient of the quantum channel in linear scale and the length of MCF, respectively. Thus the $P_{ICXT-FSRS}$ is written as
\begin{equation}
\begin{split}
\label{eq:PICXTFSRS}
P_{ICXT-FSRS}=&\int_{0}^{L} dP^{L}_{ICXT-FSRS}\\
=&\frac{\eta P_{0} exp(-\alpha_{q}L)}{2}\{ \frac{exp[(\alpha_{q}-\alpha_{c})L]-1}{\alpha_{q}-\alpha_{c}}\\
&-\frac{exp[(\alpha_{q}-\alpha_{c}-2h_{ij})L]-1}{\alpha_{q}-\alpha_{c}-2h_{ij}} \},
\end{split}
\end{equation}

Similarly, we can derive the expression of $P_{FSRS-ICXT}$. The SRS effect of a short segment ($dz$) in core-i can be expressed as
\begin{equation}
\begin{split}
\label{eq:dP2}
dP_{FSRS-ICXT}(z)=\eta P_{CS}(z)dz,
\end{split}
\end{equation}
where $P_{CS}$ is the power of classical signal at $z$ and can be expressed as \cite{ye2016design, sano2014crosstalk}
\begin{equation}
\begin{split}
\label{eq:PCS}
P_{CS}(z)=P_{0}exp(-h_{ij}z)cosh(h_{ij}z)exp(-\alpha_{c}z).
\end{split}
\end{equation}
Then the power will couple to core-j and transmit to the output of MCF. Thus the power at the output can be derived as \cite{ye2016design}
\begin{equation}
\begin{split}
\label{eq:dPL2}
dP^{L}_{FSRS-ICXT}(z)=&dP_{FSRS-ICXT}(z)ICXT(z)\\
&\times exp[-\alpha_{q}(L-z)]\\
=&dP_{FSRS-ICXT}(z)tanh(h_{ij}z)\\
&\times exp[-\alpha_{q}(L-z)],
\end{split}
\end{equation}
Thus the $P_{FSRS-ICXT}$ is written as
\begin{equation}
\begin{split}
\label{eq:PFSRSICXT}
P_{FSRS-ICXT}=&\int_{0}^{L} dP^{L}_{FSRS-ICXT}\\
=&\frac{\eta P_{0} exp(-\alpha_{q}L)}{2}\{ \frac{exp[(\alpha_{q}-\alpha_{c})L]-1}{\alpha_{q}-\alpha_{c}}\\
&-\frac{exp[(\alpha_{q}-\alpha_{c}-2h_{ij})L]-1}{\alpha_{q}-\alpha_{c}-2h_{ij}} \},
\end{split}
\end{equation}

Eventually, the total forward-ICSRS power at $z=L$ in core-j can be written as
\begin{equation}
\begin{split}
\label{eq:FICSRS2}
P_{FICSRS}=&P_{ICXT-FSRS}+P_{FSRS-ICXT}\\
=&\eta P_{0} exp(-\alpha_{q}L) \{ \frac{exp[(\alpha_{q}-\alpha_{c})L]-1}{\alpha_{q}-\alpha_{c}}\\
&-\frac{exp[(\alpha_{q}-\alpha_{c}-2h_{ij})L]-1}{\alpha_{q}-\alpha_{c}-2h_{ij}} \}.
\end{split}
\end{equation}

On the other hand, we will derive the expression of backward-ICSRS with similar method. As shown in Fig.~\ref{fig:gstd}(b), the total backward-ICSRS power is composed of two parts. The first one is the backward-SRS process of the ICXT, corresponding to the process \textcircled{3} in Fig.~\ref{fig:gstd}(b). The second one is the ICXT process of the forward-SRS, corresponding to the process \textcircled{4}. Thus the power of the backward-ICSRS can be represented as  
\begin{equation}
\begin{split}
\label{eq:BICSRS}
P_{BICSRS}=P_{ICXT-BSRS}+P_{BSRS-ICXT},
\end{split}
\end{equation}
where $P_{ICXT-BSRS}$, $P_{BSRS-ICXT}$ are the power generated by process \textcircled{3}, \textcircled{4}, respectively. The scattering power of a short segment at the input of MCF is written as
\begin{equation}
\begin{split}
\label{eq:dP0}
dP^{0}_{ICXT-BSRS}(z)=\eta P_{ICXT}(z)exp(-\alpha_{q}z)dz.
\end{split}
\end{equation}
Then the $P_{ICXT-BSRS}$ is expressed as 
\begin{equation}
\begin{split}
\label{eq:PICXTBSRS}
P_{ICXT-BSRS}=&\int_{0}^{L} dP^{0}_{ICXT-BSRS}\\
=&\frac{\eta P_{0}}{2}\{ \frac{exp[-(\alpha_{q}+\alpha_{c}+2h_{ij})L]-1}{\alpha_{q}+\alpha_{c}+2h_{ij}}\\
&-\frac{exp[-(\alpha_{q}+\alpha_{c})L]-1}{\alpha_{q}+\alpha_{c}} \}.
\end{split}
\end{equation}
Similarly, the $P_{BSRS-ICXT}$ is expressed as
\begin{equation}
\begin{split}
\label{eq:PBSRSICXT}
P_{BSRS-ICXT}=&\frac{\eta P_{0}}{2}\{ \frac{exp[-(\alpha_{q}+\alpha_{c}+2h_{ij})L]-1}{\alpha_{q}+\alpha_{c}+2h_{ij}}\\
&-\frac{exp[-(\alpha_{q}+\alpha_{c})L]-1}{\alpha_{q}+\alpha_{c}} \}.
\end{split}
\end{equation}
The total backward-ICSRS power at $z=0$ in core-j can be written as
\begin{equation}
\begin{split}
\label{eq:BICSRS2}
P_{BICSRS}=&P_{ICXT-BSRS}+P_{BSRS-ICXT}\\
=&\eta P_{0}\{ \frac{exp[-(\alpha_{q}+\alpha_{c}+2h_{ij})L]-1}{\alpha_{q}+\alpha_{c}+2h_{ij}}\\
&-\frac{exp[-(\alpha_{q}+\alpha_{c})L]-1}{\alpha_{q}+\alpha_{c}} \}.
\end{split}
\end{equation}

\section{Properties of ICSRS}
\label{properties}

\begin{figure}[!t]
\centerline{\includegraphics[width=\columnwidth]{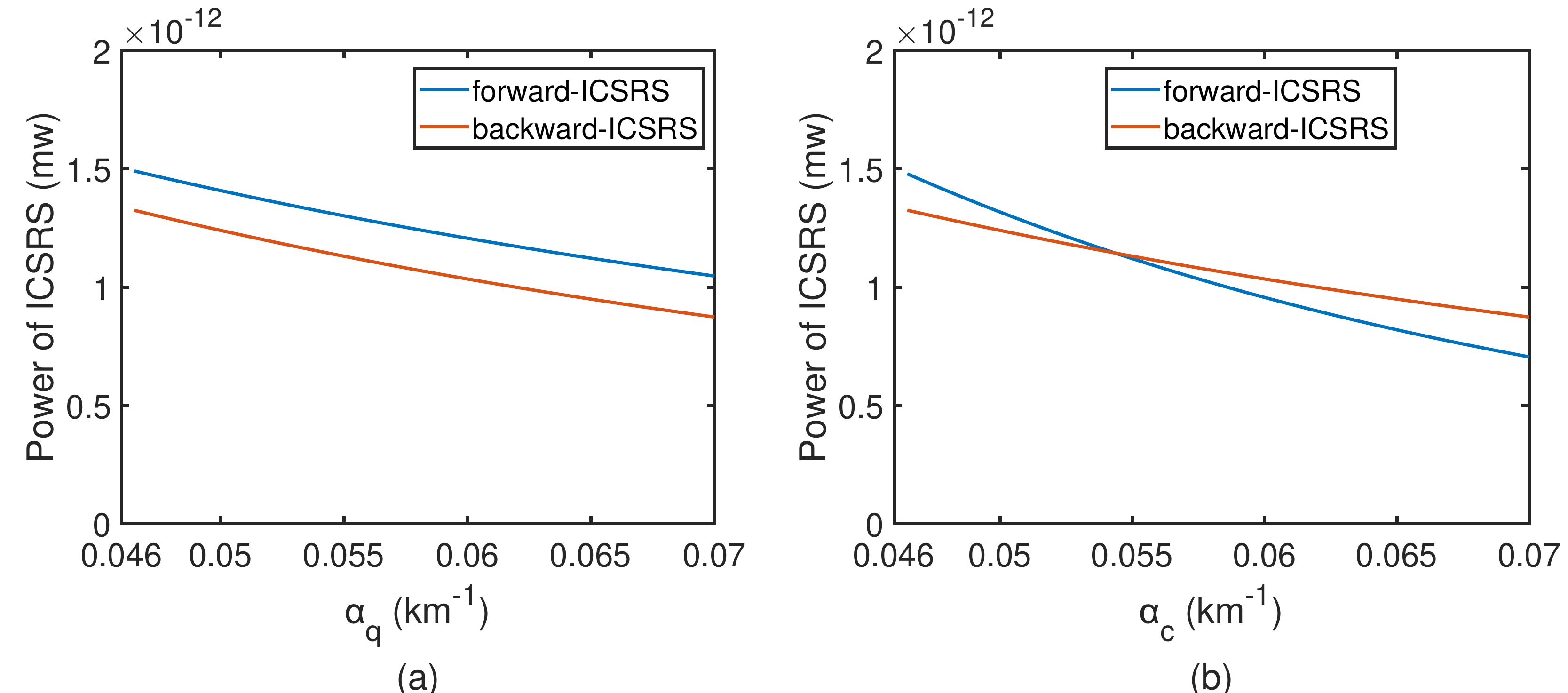}}
\caption{The power of classical signal is set to 0 dBm and the length of MCF is 50 km. $h_{ij}$ and $\eta$ are set to $10^{-6}$ $m^{-1}$ and $6\times 10^{-9}$ $(km\cdot nm)^{-1}$, respectively.}
\label{fig:FICSRS_BICSRS_deltaalpha}
\end{figure}

We will show the properties of ICSRS described by Eqs. \ref{eq:FICSRS2} and \ref{eq:BICSRS2} through the simulations. Firstly, we evaluate the impact of the attenuation coefficient on the power of ICSRS. The results are shown in Fig.~\ref{fig:FICSRS_BICSRS_deltaalpha}. In Fig.~\ref{fig:FICSRS_BICSRS_deltaalpha}(a), $\alpha_{c}$ is set to 0.046 $km^{-1}$ (0.2 dB/km) and $\alpha_{q}$ varies from 0.046 to 0.07 (about 0.3 dB/km) $km^{-1}$ ,which covers most values of attenuation coefficient in C-band. In Fig.~\ref{fig:FICSRS_BICSRS_deltaalpha}(b), $\alpha_{q}$ is set to 0.046 $km^{-1}$ and $\alpha_{c}$ varies from 0.046 to 0.07. We have to emphasize that the attenuation coefficient is wavelength dependent and $\eta$ is also wavelength dependent. However, in order to show the relationship between ICSRS power and attenuation coefficient more clearly, $\eta$ remains constant in the simulation. As can be seen from Fig.~\ref{fig:FICSRS_BICSRS_deltaalpha}, the power of ICSRS does not change dramatically when the attenuation coefficient varies. Thus attenuation coefficient is not a key parameter affecting the power of ICSRS compared with $h_{ij}$, $\eta$, etc.    

\begin{figure}[!t]
\centerline{\includegraphics[width=\columnwidth]{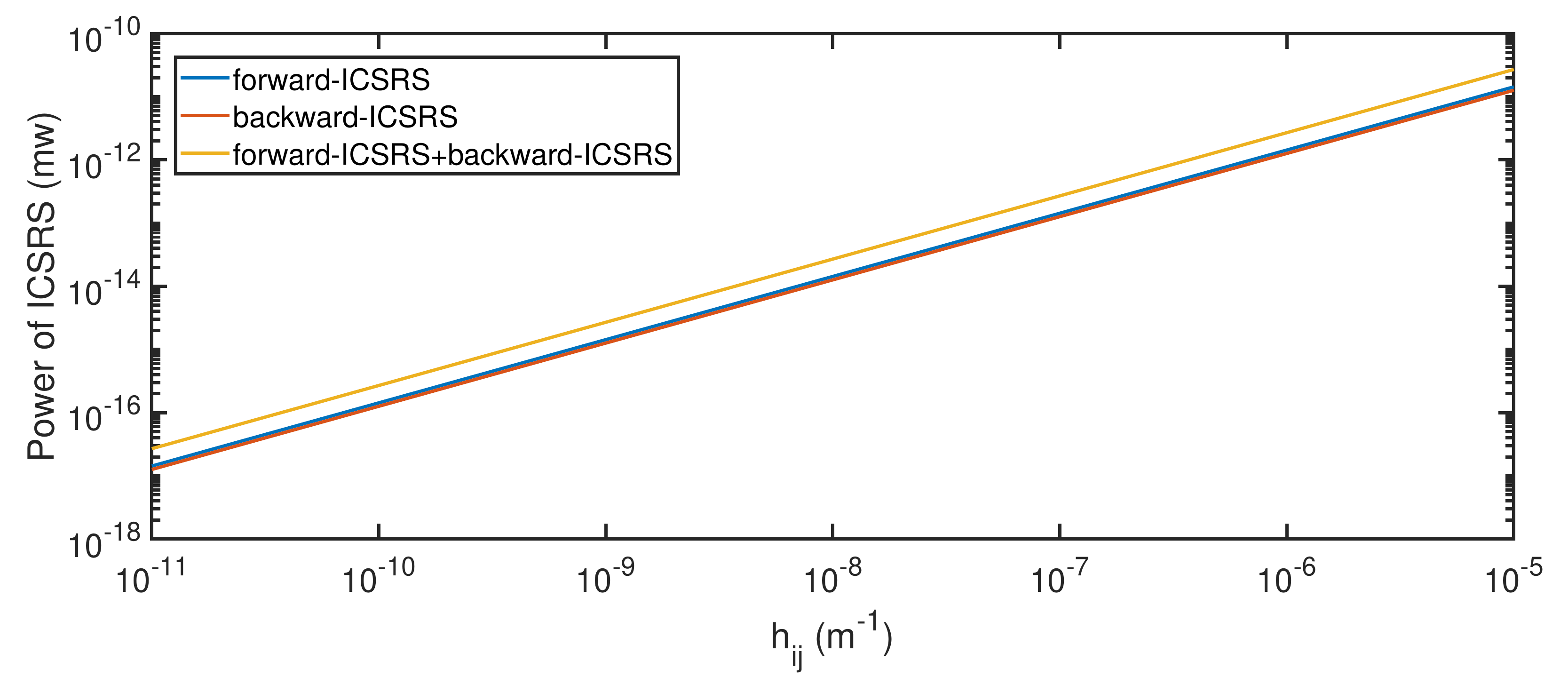}}
\caption{The power of classical signal is set to 0 dBm and the length of MCF is 50 km. $\eta$ is set to $6\times 10^{-9}$ $(km\cdot nm)^{-1}$. $\alpha_{c}$ and $\alpha_{q}$ are 0.22 and 0.21 dB/km}
\label{fig:FICSRS_BICSRS_hij}
\end{figure}

Then we evaluate the impact of $h_{ij}$ on the ICSRS power through the simulation and obtained Fig.~\ref{fig:FICSRS_BICSRS_hij}. Both the power of forward-ICSRS and that of backward-ICSRS have a approximately linear correlation with $h_{ij}$.

\begin{figure}[!t]
\centerline{\includegraphics[width=\columnwidth]{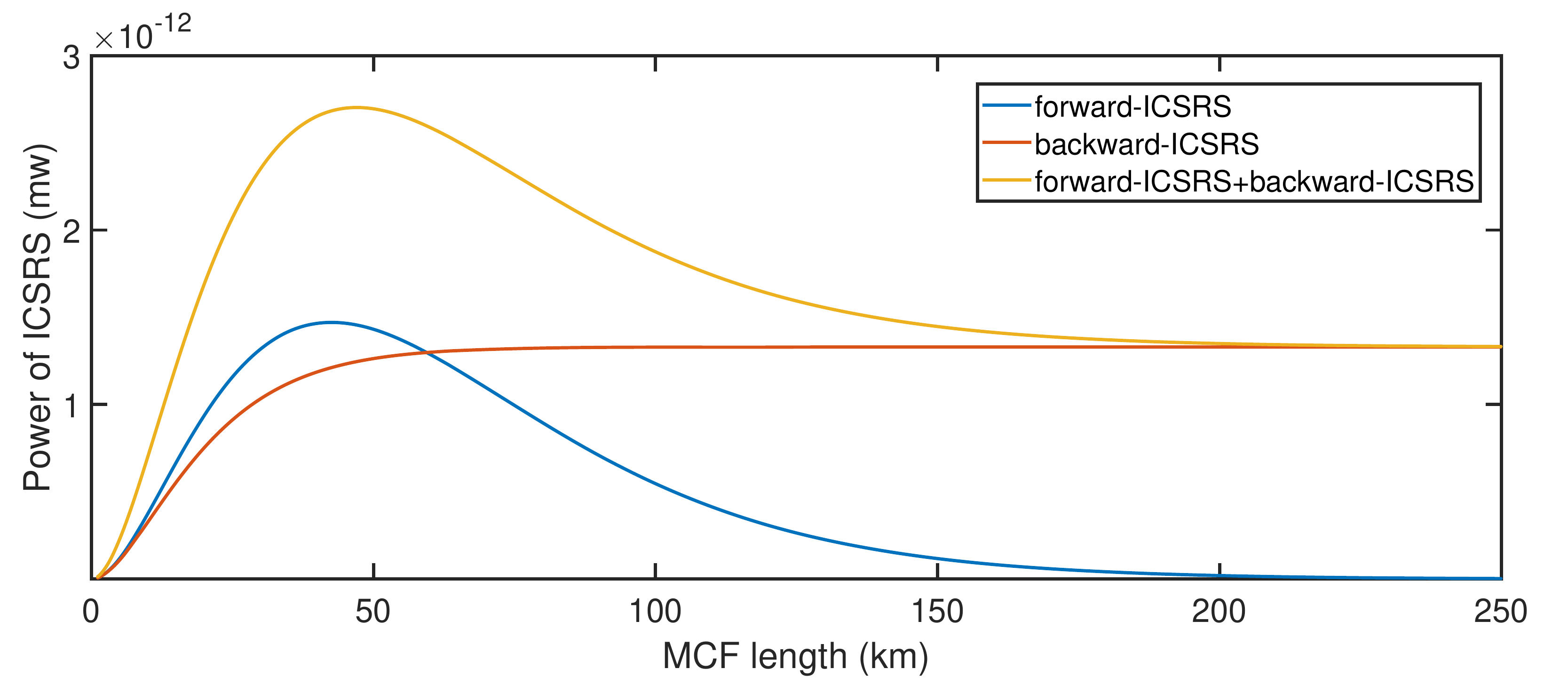}}
\caption{The power of classical signal is set to 0 dBm. $h_{ij}$ and $\eta$ are set to $10^{-6}$ $m^{-1}$ and $6\times 10^{-9}$ $(km\cdot nm)^{-1}$, respectively. $\alpha_{c}$ and $\alpha_{q}$ are 0.22 and 0.21 dB/km.}
\label{fig:FICSRS_BICSRS_L}
\end{figure}

Finally, relations between the ICSRS power and the MCF length are given in Fig.~\ref{fig:FICSRS_BICSRS_L}. The power of the forward-ICSRS reaches a maximum value at a distance of $L_{max}$ (about 40 km in Fig.~\ref{fig:FICSRS_BICSRS_L}) before it starts to decline, while that of the backward-ICSRS saturates and does not decrease with distance. In the case of forward-ICSRS, the accumulation of Raman-scattered power along the fiber is eventually outstripped by the increasing fiber attenuation, leading to a reduction of forward-ICSRS power. In contrast, backward scatter travels back to the input of MCF and is not subjected to higher loss with increasing distance. Hence, the power of backward-ICSRS never decreases but reaches saturation asymptotically.

\section{Impact of ICSRS on QKD}
\label{impact}

\begin{figure}[!t]
\centerline{\includegraphics[width=\columnwidth]{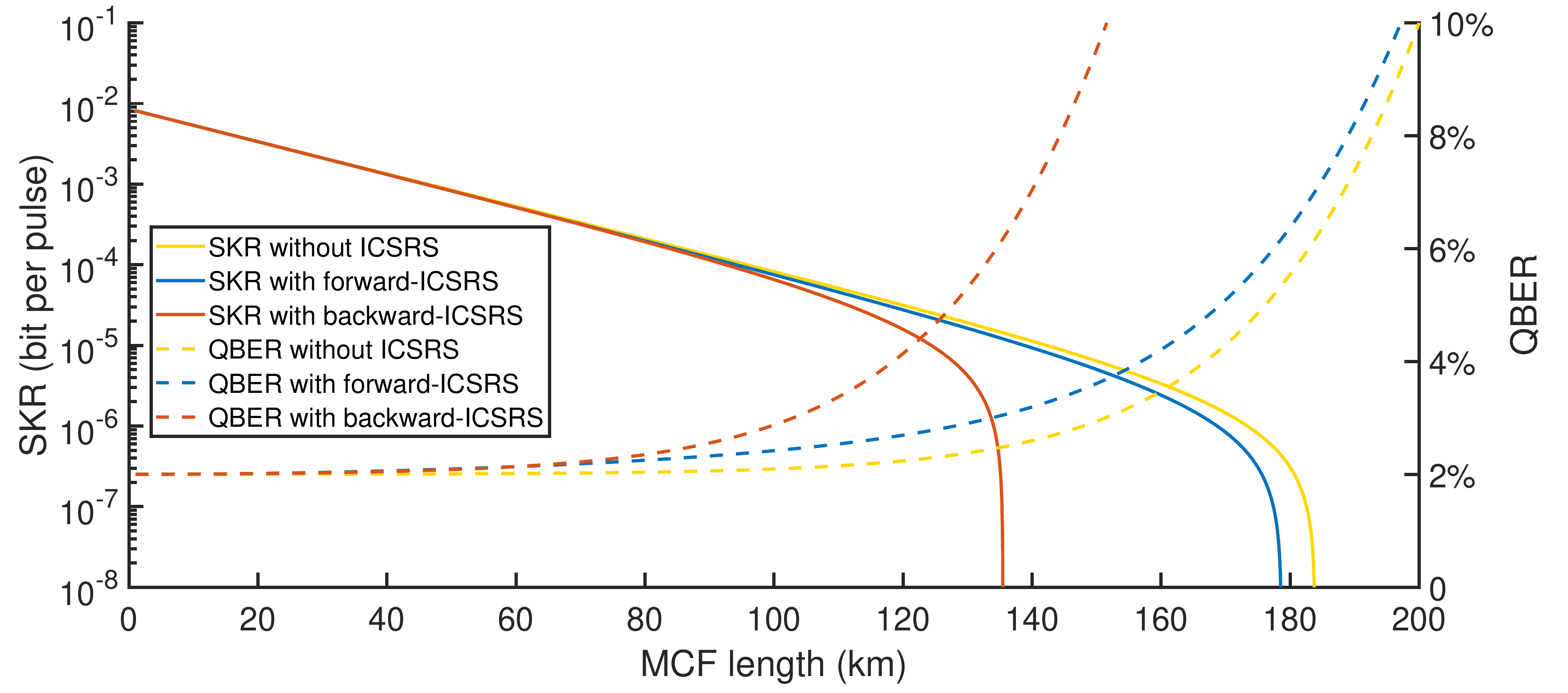}}
\caption{The power of classical signal is set to 10 dBm. $h_{ij}$ and $\eta$ are set to $10^{-6}$ $m^{-1}$ and $6\times 10^{-9}$ $(km\cdot nm)^{-1}$, respectively. $\alpha_{c}$ and $\alpha_{q}$ are 0.22 and 0.21 dB/km. The detection efficiency of the single photon detector and the dark count probability are 10\% and $1\times 10^{-6}$, respectively. The detector effective gating width and the receiving bandwith are set to 1 ns and 100 GHz, respectively.}
\label{fig:decoy_SKR_QBER_L}
\end{figure}

Firstly, we will evaluate the performance of QKD co-existing with one classical signal. The protocol used in the simulations is the BB84 protocol with decoy-state method. The secure key rate is lower bounded by \cite{lo2005decoy}
\begin{equation}
\begin{split}
\label{eq:SKR}
R=q\{-Q_{\mu}f(E_{\mu})H_{2}(E_{\mu})+Q_{1}[1-H_{2}(e_{1})]\},
\end{split}
\end{equation}
where $H_{2}$ is the binary Shannon entropy, $q$ depends on the implementation (1/2 for the BB84 protocol), $f(E_{\mu})$ is the error correction efficiency which is set to 1.15 in this paper. $Q_{\mu}$ and $E_{\mu}$ are the overall gain and the quantum bit error rate (QBER), respectively. $Q_{1}=Y_{1}\mu e^{-\mu}$ and $e_{1}=(Y_{0}/2+e_{d}t_{l})/Y_{1}$ are the gain and the error rate of a single-photon state, where $\mu$ is the average number of photons in a single pulse, $e_{d}$ represents the misalignment error, $t_{l}$ is the total transmissivity of the link. $Y_{1}=Y_{0}+t_{l}$ is the yield of a single-photon state. $Y_{0}$ is the probability of a click on the Bob’s side without having any incident photons from the transmitter. $Y_{0}$ is the yield of the vacuum state which includes the dark count of the single photon detector and
the ICSRS noise in our system. Thus, the $Y_{0}$ can be expressed as
\begin{equation}
\begin{split}
\label{eq:Y0}
Y_{0}=p_{dark}+p_{ICSRS}
\end{split}
\end{equation}
where $p_{drak}$ is the dark count rate of the single photon detector, $p_{ICSRS}$ represents the noise photon caused by the ICSRS.

The SKRs and the QBERs of QKD in different cases are shown in Fig.~\ref{fig:decoy_SKR_QBER_L}. The SKR is hardly affected by the ICSRS in short distance transmission since the SKR with ICSRS is almost the same as that without ICSRS for the transmission distance shorter than 100 km. For longer transmission distance, the SKR will be impaired by ICSRS, especially backward-ICSRS. This is because the quantum signal is greatly attenuated after the long-distance transmission and the ultra-low output power of quantum signal makes it easy to be impaired by ICSRS noise. The backward-ICSRS shows greater impairment to QKD since forward-ICSRS decreases with the transmission distance while backward-ICSRS reaches saturation asymptotically, as shown in  Fig.~\ref{fig:FICSRS_BICSRS_L}. Also, the ICSRS will limit maximum transmission distance. The maximum transmission distance is reduced by 10 km with forward-ICSRS and 50 km with backward-ICSRS. From this point of view, co-propagation in MCF for quantum signal and classical signal is a better coexistence method than counter-propagation in long distance transmission.

\begin{figure}[!t]
\centerline{\includegraphics[width=\columnwidth]{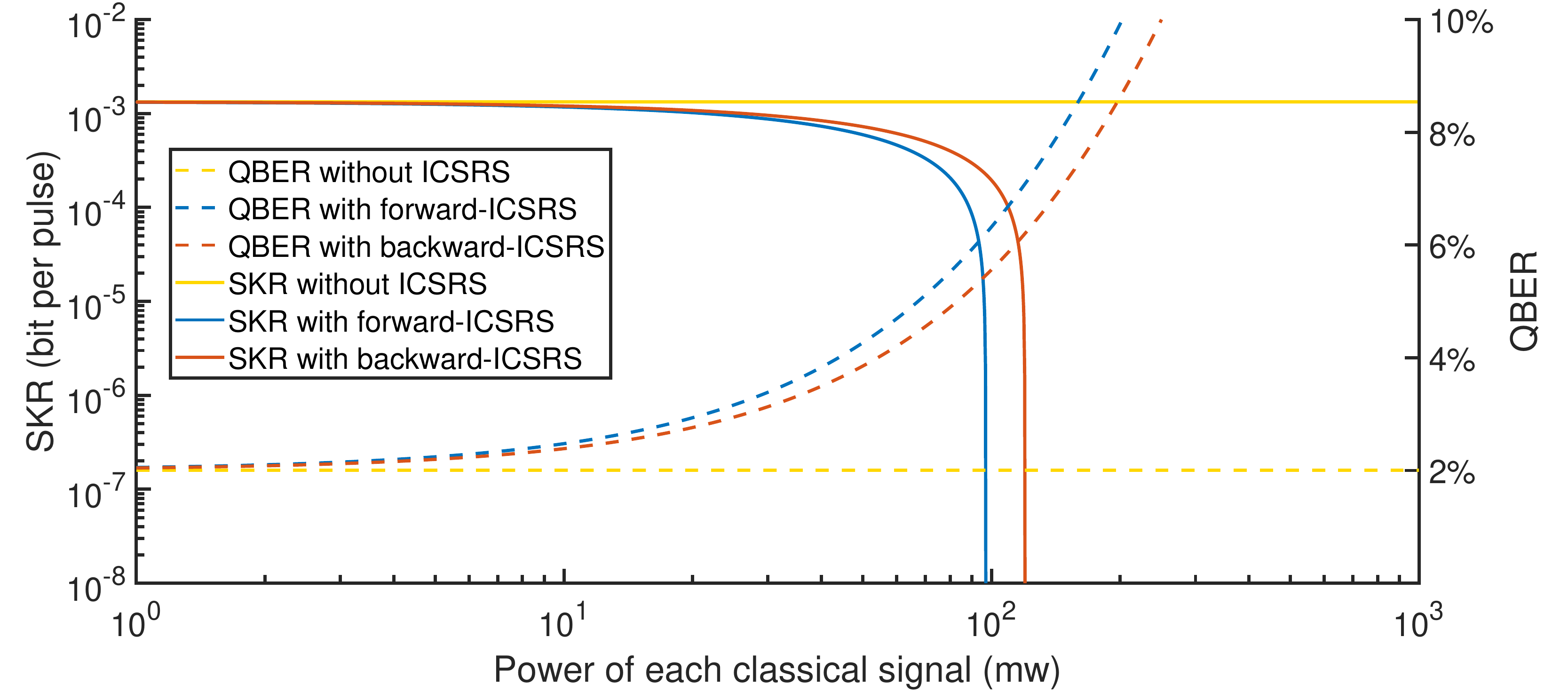}}
\caption{The MCF length is set to 40 km. $h_{ij}$ is set to $10^{-6}$ $m^{-1}$. $\alpha_{c}$ and $\alpha_{q}$ are 0.22 and 0.21 dB/km. The detection efficiency of the single photon detector and the dark count probability are 10\% and $1\times 10^{-6}$, respectively. The detector effective gating width and the receiving bandwith are set to 1 ns and 100 GHz, respectively.}
\label{fig:decoy_SKR_QBER_P}
\end{figure}

In dense wavelength division multiplexing system, many classical signals at different wavelengths will be transmitted simultanously. For certain MCF, the ICSRS in one quantum channel will be the sum of ICSRS generated from different classical channels and can be expressed as
\begin{equation}
\begin{split}
\label{eq:sumICSRS}
&P_{ICSRS}=\sum_{n=1}^NP_{FICSRS}^{n}(\alpha_{c}^{n}(\lambda_{c}^{n}),\alpha_{q}^{n}(\lambda_{q}^{n}),\eta^{n}(\lambda_{c}^{n},\lambda_{q}^{n}),P_{0}^{n})\\
&+\sum_{m=N+1}^MP_{BICSRS}^{m}(\alpha_{c}^{m}(\lambda_{c}^{m}),\alpha_{q}^{m}(\lambda_{q}^{m}),\eta^{m}(\lambda_{c}^{m},\lambda_{q}^{m}),P_{0}^{m}),
\end{split}
\end{equation}
where superscripts $n$ and $m$ means the parameter describes the $n-th$ or $m-th$ signal. $\alpha_{c}(\lambda_{c})$ and $\alpha_{q}(\lambda_{q})$ do not change dramaticly with the wavelength in C-band (between 0.048 and 0.053 $km^{-1}$ \cite{kawahara2011effect}) and the power of ICSRS will not vary greatly with such attenuation coefficient which is shown in Sec. \ref{properties}. Thus, for the sake of simplicity, the attenuation coefficient is assumed constant in the simulation. Then the power of the ICSRS can be approximated as
\begin{equation}
\begin{split}
\label{eq:approsumICSRS}
P_{ICSRS}\approx &G\sum_{n=1}^N\eta^{n}(\lambda_{c}^{n},\lambda_{q}^{n})\cdot P_{0}^{n}\\
&+F\sum_{m=N+1}^M\eta^{m}(\lambda_{c}^{m},\lambda_{q}^{m})\cdot P_{0}^{m},
\end{split}
\end{equation}
where 
\begin{equation}
\begin{split}
\label{eq:G}
G=&exp(-\alpha_{q}L) \{ \frac{exp[(\alpha_{q}-\alpha_{c})L]-1}{\alpha_{q}-\alpha_{c}}\\
&-\frac{exp[(\alpha_{q}-\alpha_{c}-2h_{ij})L]-1}{\alpha_{q}-\alpha_{c}-2h_{ij}} \},
\end{split}
\end{equation}
\begin{equation}
\begin{split}
\label{eq:F}
F=&\frac{exp[-(\alpha_{q}+\alpha_{c}+2h_{ij})L]-1}{\alpha_{q}+\alpha_{c}+2h_{ij}}\\
&-\frac{exp[-(\alpha_{q}+\alpha_{c})L]-1}{\alpha_{q}+\alpha_{c}}.
\end{split}
\end{equation}

The feasibility of the co-existence of QKD and classical signals over the range of metropolitan area networks (typically 20-40 km) is validated in Fig.~\ref{fig:decoy_SKR_QBER_P}. Co-existence of 16 classical signals with equal power and one quantum signal is shown in the simulation. The frequencies of 8 classical signals are set lower than quantum signal while those of the other 8 classical signals are set higher than quantum signal. The frequency of quantum signal is 193.5 THz and the frequency spacing of each signal is 200 GHz. Then we can get the value of $\eta$ between quantum signal and each classical signal in Fig. 1 in \cite{eraerds2010quantum}. In order to show the potential to support QKD along with high bandwidth data transport, the power of each classical signal is increased in the simulation. As can be seen in Fig.~\ref{fig:decoy_SKR_QBER_P}, the QBER and SKR remain almost constant when the power of classical signal is not so high. Then, the SKR will decrease with the classical signal power of 10 mw and it will decrease dramatically when the power of classical signal increases. No secure keys can be generated by the QKD system with the classical signal power of about 100 mw.

\section{Compared with SRS in single core fiber}
\label{compare}

The forward-SRS generated in single core fiber can be expressed as  
\begin{equation}
\begin{split}
\label{eq:FSRS}
P_{FSRS}=\frac{\eta P_{0}}{\alpha_{q}-\alpha_{c}}[exp(-\alpha_{c}L)-exp(\alpha_{q}L)]
\end{split}
\end{equation}
and the BSRS is expressed as
\begin{equation}
\begin{split}
\label{eq:BSRS}
P_{BSRS}=\frac{\eta P_{0}}{\alpha_{q}+\alpha_{c}}[exp(\alpha_{c}L)-exp(\alpha_{q}L)]exp(-\alpha_{c}L).
\end{split}
\end{equation}

\begin{figure}[!t]
\centerline{\includegraphics[width=\columnwidth]{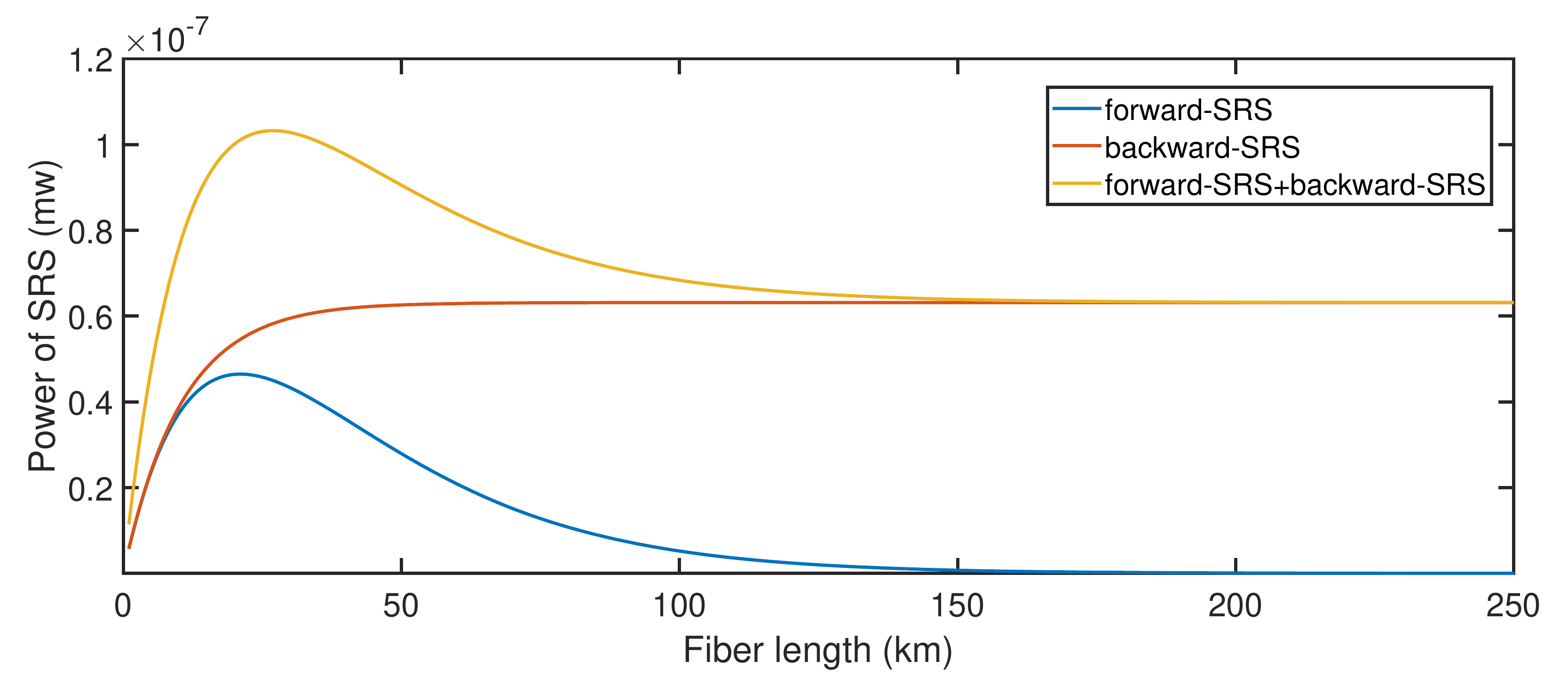}}
\caption{Power of SRS as a function of fiber length.}
\label{fig:compareSRS}
\end{figure}

The power of SRS with the fiber length is plotted in Fig.~\ref{fig:compareSRS}. Compared with the power of ICSRS in Fig.~\ref{fig:FICSRS_BICSRS_L}, the power of SRS shows similar trend. The power of backward-SRS reaches saturation asymptotically with the fiber length and that of forward-SRS will decline after reaching a maximum value at a distance of $L_{max}^{\prime}$ (about 20 km in Fig.~\ref{fig:compareSRS})

In short distance, the power of forward-ICSRS is higher than that of backward-ICSRS while the power of forward-SRS is lower than that of backward-SRS. This is because the power of source classical signal varies differently with the distance in MCF and single core fiber. For forward-SRS, the peak power is at the distance of $L_{max}^{\prime}$ which is expressed
\begin{equation}
\begin{split}
\label{eq:Lmaxprime}
L_{max}^{\prime}=\frac{ln(\alpha_{q}/\alpha_{c})}{\alpha_{q}-\alpha_{c}}.
\end{split}
\end{equation}
When the classical signal and quantum signal are in C-band, the $L_{max}^{\prime}$ is around 25 km. For forward-ICSRS, we obtain the value of $L_{max}$ through the method of traversing. The result shows that the $L_{max}$ is between 30 and 45 km when the classical signal and quantum signal are set in C-band and $h_{ij}$ varies between $10^{-11}$ and $10^{-6}$ $m^{-1}$. Thus, the $L_{max}$ is always longer than $L_{max}^{\prime}$.

\section{Conclusion}
In this paper, mathematical expression for ICSRS is derived for the first time, based on the effect of ICXT in MCF. Then the properties of forward-ICSRS and backward-ICSRS are studied through numerical simulations. The results show that both the power of forward-ICSRS and that of backward-ICSRS have approximately linear correlation with power coupling coefficient. The power of the forward-ICSRS reaches a maximum value with the transmission distance and then it starts to decline, while that of the backward-ICSRS saturates and does not decrease with distance. Then the impact of ICSRS on QKD is evaluated. It shows that both forward-ICSRS and backward-ICSRS will reduce the maximum transmission distance of QKD and backward-ICSRS has more impairment to QKD. Over the range of metropolitan area networks, the quantum signal is affected only when the power of classical signal is very high in the dense wavelength division multiplexing system. Finally, SRS generated in single core fiber and the ICSRS generated in MCF are compared. It is revealed that the power of ICSRS has similar properties with SRS. However, the transmission distance of forward-ICSRS where it reaches the power peak is longer than that of forward-SRS.

% or


%\appendices
%section{Proof of the First Zonklar Equation}

%\section{}

%\section*{Acknowledgment}


\ifCLASSOPTIONcaptionsoff
  \newpage
\fi



%\bibliographystyle{IEEEtran}
%\begin{thebibliography}{1}


%\end{thebibliography}
\bibliographystyle{IEEEtran}
\bibliography{IEEEabrv,bare_jrnl}

\end{document}